\documentclass{article}
\usepackage[a4paper, margin={3.0cm}]{geometry}
\usepackage{graphicx}
\usepackage{amsmath}
\usepackage{amsfonts}
\usepackage{amssymb}
\usepackage{mathtools}
\usepackage{enumitem}
\usepackage{hyperref}
\usepackage{algorithmic}
\usepackage{algorithm}
\usepackage{listings}
\usepackage{color}

\newlength{\topFigureVerticalSpace}
\newlength{\bottomFigureVerticalSpace}
\newcommand{\tfvspace}{\vspace*{\topFigureVerticalSpace}}
\newcommand{\bfvspace}{\vspace*{\bottomFigureVerticalSpace}}

\newcommand{\magma}{\textsc{Magma}}
\newcommand{\savedoneg}{\textsc{Saved\_1\_$\Gamma$}}
\newcommand{\savedtwog}{\textsc{Saved\_2\_$\Gamma$}}
\newcommand{\savedisom}{\textsc{Saved\_isometry}}

\newcommand{\modified}{\textsc{modified\_Brouwer\_Zimmermann}}
\newcommand{\modifieds}{\textsc{modified\_Brouwer\_Zimmermann$_s$}}

\newcommand{\diagFdos}{\textsc{Diagonalization\_over\_$\mathbb{F}_2$}}
\newcommand{\diagFcuatro}{\textsc{Diagonalization\_over\_$\mathbb{F}_4$}}

\newcommand{\convertFTwo}{%
  \textsc{Transform\_Matrix\_$\mathbb{F}_4$\_$\mathbb{F}_2$}%
}

\newcommand{\convertFFour}{%
  \textsc{Transform\_Matrix\_$\mathbb{F}_2$\_$\mathbb{F}_4$}%
}

\newcommand{\iso}{\textsc{Isometry\_Transformation}}

\newcommand{\wt}{\textrm{wt}}
\newcommand{\np}{n_p}
\newcommand{\Pnp}{n_{pp}}

\makeatletter
\renewcommand*{\@fnsymbol}[1]{\@arabic{#1}}
\makeatother

\title{Fast Algorithms and Implementations
       for Computing the Minimum Distance of Quantum Codes}

\author{%
Fernando Hernando%
\thanks{%
  Depto.~de Matem\'aticas,
  Universidad Jaume I,
  12.071--Castell\'on, Spain.
  \texttt{carrillf@mat.uji.es}
}
\and
Gregorio Quintana-Ort\'{\i}%
\thanks{%
  \raggedright Depto.~de Ingenier\'{\i}a y Ciencia de Computadores,
  Universidad Jaume I,
  12.071--Castell\'on, Spain.
  \texttt{gquintan@uji.es}
} 
\and
Markus Grassl%
\thanks{%
  \raggedright International Centre for Theory of Quantum Technologies (ICTQT),
  University  of Gdansk, Gda\'nsk, Poland.
  \texttt{markus.grassl@ug.edu.pl}
}
}

\date{\today}

\begin{document}
% =============================================================================

\maketitle

\begin{abstract}
The distance of a stabilizer quantum code 
is a very important feature since it determines the number of errors
that can be detected and corrected.
We present three new fast algorithms and implementations
for computing the symplectic distance of the associated classical code.
Our new algorithms are based on the Brouwer-Zimmermann algorithm.
Our experimental study shows that these new implementations are much faster 
than current state-of-the-art licensed implementations on 
single-core processors,
multicore processors, and
shared-memory multiprocessors.
In the most computationally-demanding cases,
the performance gain in the computational time can be 
larger than one order of magnitude.
The experimental study also shows a good scalability on 
shared-memory parallel architectures.
\end{abstract}

% =============================================================================
\section{Introduction}
% =============================================================================

Quantum computing can be considered 
as a paradigm shift in computational theory,
poised to redefine the boundaries of theory and practice of computing.
Departing from the binary system of classical computing, quantum computers
leverage the principles of quantum mechanics to operate with quantum bits,
or qubits.
Unlike classical bits, which can be either $0$ or $1$, qubits can exist in
multiple states simultaneously 
due to the phenomenon of superposition~\cite{nielsen_quantum_2010}.
Additionally, qubits can be entangled, a property that allows for
instantaneous correlation between them, 
no matter the distance~\cite{nielsen_quantum_2010}.

This fundamental difference enables quantum computers to perform certain
calculations exponentially faster, promising groundbreaking advancements
across various domains, including cryptography, optimization, and material
science.
In cryptography, quantum computers have the potential to break widely-used
encryption methods, requiring the development of quantum-resistant
algorithms~\cite{shor_algorithm_1994}.
In optimization, they can solve complex problems in logistics, finance, and
artificial intelligence more efficiently 
than classical computers~\cite{farhi_quantum_2014}.
In material science, quantum simulations could lead to the discovery of new
materials and drugs 
by accurately modeling molecular interactions~\cite{mcardle_quantum_2020}.

Quantum error-correcting codes (QECCs) 
are a key component to the practical viability of quantum computing, 
addressing the inherent fragility
of quantum systems to noise and decoherence.
In essence, quantum error correction addresses the fundamental challenge of
preserving quantum coherence in the presence of environmental disturbances.
Without error correction, quantum information stored in qubits would
quickly degrade, rendering quantum computations unreliable.
By leveraging the principles of quantum mechanics, quantum error-correcting
codes can detect and correct errors without directly measuring the qubits,
thus preserving the delicate quantum superposition,
which is essential for quantum computing~\cite{gottesman_stabilizer_1997}.

The development of fault-tolerant quantum computation, made possible by
quantum error correction, is crucial in scaling quantum algorithms to
tackle real-world challenges effectively.
As quantum computers continue to evolve, the synergy between quantum
computing and quantum error correction underscores the transformative
potential of quantum technologies in revolutionizing computation and
information processing.

Most of the quantum error-correcting codes discussed in the
literature are stabilizer codes, 
whose minimum distance serves as 
a critical measure of error detection and correction capabilities.
Stabilizer codes, such as those introduced by Daniel Gottesman and the
quantum error correction codes over $GF(4)$ pioneered by Calderbank and Shor,
encode qubits into highly entangled states, thereby safeguarding quantum
information against disruptions.
These codes play a vital role in preserving quantum coherence and
maintaining the integrity of quantum information 
during computation~\cite{gottesman_stabilizer_1997,calderbank_quantum_1996,ketkar_nonbinary_2010,nielsen_quantum_2010}.

The aim of our work is to accelerate the computation of the minimum distance 
of a random stabilizer code, 
a task essential for assessing its error-correcting capabilities.
By determining the minimum distance, 
insights into the code's ability to detect and correct errors is gained,
which is key to the practical viability of quantum computing.

This analysis contributes to the ongoing efforts in quantum error
correction, paving the way for the development of fault-tolerant quantum
computation.
Such advancements are fundamental for scaling quantum algorithms to
effectively tackle real-world challenges across various domains, 
showcases the revolutionary impact of quantum technologies 
on computation and data processing.

Stabilizer codes, essential in quantum coding, 
emerge when the associated classical stabilizer code is a subset of its symplectic dual, 
leading to quantum codes with specific properties.
Our study focuses on computing the minimum distance of 
randomly-generated quantum stabilizer codes for qubits
by analyzing the symplectic weight of linear combinations 
of rows of the normalizer matrix.
To achieve this, 
we have developed three new algorithms and implementations.
Each one employs unique approaches for matrix manipulation and 
diagonalization to compute the symplectic minimum distance.

Our experimental analysis assesses these three new implementations
and compares them to a state-of-the-art licensed software (\magma{}).
The performance assessments encompassed matrices of different 
sizes and complexities.
Our implementations consistently outperformed \magma{},
being faster for most of the codes assessed.
Particularly, the speedups were remarkable:
Our new implementations were up to $40$ times as fast as \magma{}
in the most-demanding cases.

The rest of the paper is organized as follows:
Section~\ref{algorithms} describes the new algorithms and implementations 
developed in our work.
Section~\ref{performance} contains a comparative analysis of
the implementations of our new algorithms.
Finally,
Section~\ref{conclusions} enumerates the conclusions of our work.

% =============================================================================
\section{Algorithms and implementations}
\label{algorithms}
% =============================================================================

The symplectic weight of any vector $(a,b)\in \mathbb{F}_q^{2n}$, 
with $a,b\in\mathbb{F}_q^n$, is defined as follows:
$$
  \wt{}_s(a,b)=\#\{i\mid a_i\ne 0 \ \text{or}\  b_i\ne 0\}.
$$

For any set $C\subset \mathbb{F}_q^{2n}$, 
the symplectic weight is defined
as the minimum of all such weights within $C$:
$$
  \wt{}_s(C)=\min\{\wt{}_s(a,b)\mid (a,b)\in C\setminus\{(0,0)\}\}.
$$

Moreover, the symplectic inner product over $\mathbb{F}_q^{2n}$ is defined as:
$$
  (a,b) \cdot_s (c,d)=a\cdot d - b\cdot c,
$$
where $a\cdot d$ and $b\cdot c$ denote the Euclidean inner products.

Given an $\mathbb{F}_q$-linear subspace $C\subset \mathbb{F}_q^{2n}$, the
symplectic dual is defined as:
$$
  C^{\perp_s}=\{(a,b)\in\mathbb{F}_q^{2n}\mid  (a,b) \cdot_s (c,d)=0, \quad
  \forall (c,d)\in C\}.
$$

Stabilizer codes are constructed as follows: 
Given $C\subset \mathbb{F}_q^{2n}$, 
an $\mathbb{F}_q$-linear subspace 
with parameters $[2n,n-k,d]_q$ such that $C\subset C^{\perp_s}$, 
then there exists a quantum code $Q\subset(\mathbb{C}^q)^{\otimes n}$ 
with parameters $[\![n,k,d(Q)]\!]_q$, 
where $d(Q)=\wt{}_s(C^{\perp_s}\setminus C)$ for $k>1$, and $d(Q)=\wt{}_s(C)$ for $k=0$.
The code $C$ is referred to as the \emph{classical stabilizer code} associated to $Q$, 
and its symplectic dual $C^{\perp_s}$ is referred to 
as the \emph{classical normalizer code} associated to $Q$.

The objective of this study is to compute the minimum distance $d(Q)$ of a
randomly generated quantum stabilizer code $Q$.
To achieve this, one must examine a generator matrix of the classicial normalizer code associated to $Q$, which is an $(n+k)\times 2n$ matrix $A$ over $\mathbb{F}_q^{2n}$.
We refer to $A$ as the \emph{normalizer matrix} of $Q$.
Consequently, the symplectic weight of any $\mathbb{F}_q$-linear
combination of rows of $A$ becomes the focal point of research.

Notice that in \magma{} the normalizer matrix $A$ 
in so-called \emph{extended form} is obtained from $Q$ as:
\begin{lstlisting}
  A := NormalizerMatrix( Q: ExtendedFormat := true );
\end{lstlisting}

We focus on the particular case where $q=2$, i.e., so-called qubit codes.
Therefore, we study the symplectic weight of the sum over any subset of rows of $A$,
with the restriction that the resulting codeword must be 
in $C^{\perp_s}\setminus C$ when $k>0$.

We have implemented three different algorithms 
for computing the minimum distance of a random qubit stabilizer code,
namely: \savedoneg, \savedtwog, and \savedisom.
The details of these algorithms and implementations will be described next.

% -----------------------------------------------------------------------------
\subsection{Modified Brouwer-Zimmerman Algorithm}
% -----------------------------------------------------------------------------

Determining the minimum weight of a random linear code $C$ 
requires finding the smallest non-zero Hamming weight among all codewords in $C$,
which can be defined as \( \min\{\wt{}(c) \mid c \in C\setminus\{0\}\} \).
When the code operates over \( \mathbb{F}_q \) and 
has a dimension of $k$, it has $q^k$ codewords.
Obviously, computing the minimum of the weight of all those codewords
becomes unfeasible even for medium and large values of $q$ and $k$.

The fastest general algorithm for computing the minimum distance 
of a random linear code is 
the so-called Brouwer-Zimmermann algorithm~\cite{Zim},
which is described in detail by Grassl~\cite{Grassl}.
There is an implementation of this algorithm
in the closed-source \magma{}~\cite{Magma} over any finite field, 
whereas there is an open-source implementation
in GAP (concretely, in the Guava package)~\cite{Guava,GAP} 
over the fields $\mathbb{F}_2$ and $\mathbb{F}_3$.

To address the challenge of computing the minimum weight, the Brouwer-Zimmerman algorithm 
introduces an upper bound $U$ and a lower bound $L$.
If \( L \geq U \), then the true minimum weight is $U$.
The upper bound $U$ is updated
whenever a new codeword $c$ is discovered with a weight \( \wt{}(c) \) 
lower than $U$.
On the other hand, the lower bound $L$ is derived from using possibly 
more than one systematic generator matrix $\Gamma_i$ for the code $C$.
Considering linear combinations of $w$ rows, the resulting codeword has weight at least $w$ in each of the information sets.
The lower bound $L$ can be increased every time a new enumeration 
of linear combinations of rows of 
the generator matrices $\Gamma_i$ initiates with a different cardinality.
This approach streamlines the process, significantly reducing the computational times.

The Modified Brouwer-Zimmerman algorithm additionally checks whether a discovered codeword $c$ with weight lower than $ U$ is also in \( C^{\perp_s} \setminus C \) before updating  $U$. In the exceptional case where $ k = 0 $, the code is self-dual, and the check to see whether it is in \( C^{\perp_s} \setminus C \) does not apply. In this situation, we always update $U$ with the new codeword's weight.
We denote the case where the algorithm assesses the minimum Hamming
distance as \modified, and the case where it assesses the symplectic minimum distance 
as \modifieds{}.

A fast implementation of the Brouwer-Zimmermann algorithm for 
both single-core architectures 
and shared-memory parallel architectures 
(multicore and multiprocessor architectures)
was recently introduced~\cite{HIQ}.
This work offered performances about two or three times as fast 
as those of \magma{} and Guava (GAP) over $\mathbb{F}_2$
by saving and reusing the additions of combinations.
A novel implementation for multicomputers and 
distributed-memory architectures was also recently introduced~\cite{HIQ2}.
It allows the use of thousands of cores for computing the distance,
thus notably reducing the total computational time from days to seconds.

We have taken advantage of these algorithms and implementations
for single-core architectures and shared-memory parallel architectures,
based on the idea of saving and reusing the additions of combinations,
to develop new fast implementations 
for computing the minimum distance of quantum stabilizer codes.
Next, we describe our new three algorithms and implementations 
for computing the minimum distance of quantum stabilizer codes.

% -----------------------------------------------------------------------------
\subsection{Algorithm \savedoneg}
% -----------------------------------------------------------------------------

Given a matrix $A$ of dimension $(n+k) \times 2n$, where $0\le k\le n$,
we must ensure that the $n+k$ rows are linearly independent over
$\mathbb{F}_2$.
Although standard diagonalization algorithms can be applied, 
an additional constraint is imposed: 
any column permutation applied within the first half of the columns
must be correspondingly applied within the second half (and vice versa).
Additionally, one may swap the $i$-th column in the first half with the $i$-th column in the second half.
This restriction arises from the symplectic weight's consideration 
of mirrored coordinates, 
prohibiting the permutation of one without its corresponding reflection.

This process results in a matrix of the form:
\[
  B = \left( 
        \begin{array}{@{}c|c@{}}
          I_n & M_1 \\
          0   & M_2
        \end{array}
      \right)
\]
where $I_n$ is the identity matrix of dimension $n$,
$M_1$ is an $n\times n$ matrix, and 
$M_2$ is a $k\times n$ matrix in row-echelon form.

Next, for $i=1,\ldots,k$, determine the column $j_i$ in the matrix $M_2$ such that the position $(i,j_i)=1$ and all elements to the left and all elements below are zero.
The row $j_i$ of the matrix $B$ is added to the $i$-th row for $M_2$, and the sum is appended to $B$ as a new row.
Let $\widetilde{B}$ be the resulting modified matrix
and let this process be denoted as \diagFdos, 
then $\widetilde{B}=\diagFdos(A)$.

Afterwards, the \modifieds{} is employed, assessing the symplectic
weight instead of the Hamming weight.

The rationale for considering the matrix $\widetilde{B}$ 
instead of the matrix $A$ is illustrated with the following example.
Assume $A$ is defined as follows:
\[
  A=\left(\begin{array}{@{}cc|cc@{}}
    1 & 0 & 0 & 1  \\
    0 & 1 & 1 & 0  \\
    0 & 0 & 1 & 1
  \end{array}\right) .
\]
The matrix $A$ is already in echelon form, so $A=B$. The matrix $\widetilde{B}$ is given by
\[
  \widetilde{B}=\left(\begin{array}{@{}cc|cc@{}}
    1 & 0 & 0 & 1  \\
    0 & 1 & 1 & 0  \\
    0 & 0 & 1 & 1 \\
    1 & 0 & 1 & 0
  \end{array}\right) .
\]

When \modifieds($A$) is applied, it first considers the symplectic weight of each row of $A$ and finds the upper bound $U=2$.
Then, before considering the sum of any two rows, the lower bound $L$ is set to $2$. Hence $L\ge U$, and the result is a symplectic minimum distance of $2$.
In contrast, applying \modifieds($\widetilde{B}$) results in the correct symplectic minimum distance of $1$.
This example demonstrates the importance of considering $\widetilde{B}$ to achieve the true symplectic minimum distance.

\begin{algorithm}[ht!]
  \caption{\ensuremath{\mbox{\sc \savedoneg}}}
  \label{alg:saved1g}
  \begin{algorithmic}[1]
    \REQUIRE A normalizer matrix $A$ of size $(n+k)\times 2n$ 
             of a quantum stabilizer code $Q$.
    \ENSURE  The minimum weight of $Q$.
    \medskip
    \STATE \textbf{Beginning of Algorithm}
    \STATE $\widetilde{B}$ := \diagFdos(A);
    \STATE d := \modifieds($\widetilde{B}$);
    \RETURN d;
    \STATE \textbf{End of Algorithm}
  \end{algorithmic}
\end{algorithm}

% -----------------------------------------------------------------------------
\subsection{Algorithm \savedtwog}
% -----------------------------------------------------------------------------

This algorithm works on a matrix over \( \mathbb{F}_4 \)
resulting from a modification of \( A \).
Let assume that \( \{1,\alpha\} \) forms a basis of \( \mathbb{F}_4 \) 
over \( \mathbb{F}_2 \).
Then, a new matrix over \( \mathbb{F}_4 \), denoted as \( A_4 \),
is constructed as follows: 
The rows of \( A_4 \) are formed by \( a+\alpha b \), 
where \( (a, b) \) represents a row of \( A \).
It is worth noting that the linear code over \( \mathbb{F}_2 \) with
generator matrix \( A \) is isometric to the additive code with generator
matrix \( A_4 \). The symplectic weight for the code over $\mathbb{F}_2$ of length $2n$ corresponds to the Hamming weight for the code over $\mathbb{F}_4$ of length $n$.
Here, by ``additive,'' 
we refer to linearity over \( \mathbb{F}_2 \) rather than over \( \mathbb{F}_4 \), 
therefore only additions between rows are allowed.

Diagonalization over $A_4$ is computed,
noting its linearity over $\mathbb{F}_2$.
This restricts us to considering row additions exclusively, as
multiplications by $\alpha$ or $\alpha^2$ are prohibited.
Thus, one can, for example, nullify $\alpha^2$ within a given row only by adding another row with $\alpha^2$ at the same position, or by adding
two other rows containing $\alpha$ and $1$, respectively, at the same
position.

Subsequently, columns and rows are permuted.
The current column can be swapped only with one of the next columns.
Analogously, 
the current row can be swapped only with one of the next rows.
The current row is also use to ``clear'' the entries in the same column above and below. Here ``clear'' refers to making all the other elements either zero or equal to a non-zero element of $\mathbb{F}_4$ different from the current pivot element.

Columns containing two different non-zero symbols at the desired positions are prioritized, resorting to row permutation if necessary.  When there are two different non-zero symbols, all the other entries in the same column can be cleared.
However, it is possible that among the remaining columns, 
a column with two different non-zero symbols cannot be found at the desired positions.
Consequently, the last columns may have more than two non-zero entries.
Thus, at the conclusion of the diagonalization, 
for example, a matrix of the following form is obtained:

\[
A_4=\left(\begin{matrix}
  1      & 0        & 0        & \star  & * & * \\
  \alpha & 0        & 0        & \star  & * & * \\
  0      & 1        & 0        & \star  & * & * \\
  0      & \alpha^2 & 0        & \star  & * & * \\
  0      & 0        & \alpha   & \star      & * & * \\
  0      & 0        & \alpha^2 & \star  & * & * \\
  0      & 0        & 0        & \alpha & * & * \\
\end{matrix}\right)
\]

In the example, the entries marked by $\star$ in the fourth column are either $0$ or $1$.   

Next, the following algorithm is implemented, processing the columns from left to right:
Whenever a column contains exactly two non-zero elements that are different, if the corresponding rows have not yet been processed, their sum is appended as a new row to $A_4$.
The resulting matrix at the end of this process is denoted as $B_4$, denoted by
$B_4$ := \diagFcuatro($A_4$).

Know we explain the reason for these extra rows and why the algorithm works. 
Whenever we add a new row to $A_4$, the corresponding column in the final matrix $B_4$ will have exactly three non-zero entries 
$1$, $\alpha$, and $\alpha^2$. The corresponding rows form a \emph{package}, corresponding to an element of an information set in the linear case.  Rows which have only a single pivot element, like the last row of $A_4$ in the example, form a package with a single element.  For a package with three element, the sum of any pair of rows equals the other row, and hence we do not need to consider sums of different rows from the same package.  For packages with a single row, the element in the corresponding column cannot be cancelled by the entries in other rows.
Therefore, if we consider the sum of $g$ rows of the final matrix $B_4$, we can ensure that the Hamming weight is at least $g$ if
each row comes from a different packet. In other words, the weight is higher than or equal to  the number of packets involved in the sum. 
So, after enumerating all the  linear combinations of up to $g$ rows of $B_4$, where each row comes from a different package, we can obviously guarantee that $L>g$.

This is illustrated with with the following example.
Assume $A_4$ is defined as follows:
\[
A_4=\left(\begin{matrix}
  1      & 1        & 1        & 1 \\
  \alpha & 0   & \alpha   & \alpha  \\
  0      & \alpha^2 & \alpha^2 & \alpha^2
\end{matrix}\right)
\]
The matrix $B_4$ is given by
\[
B_4=\left(\begin{matrix}
  1        & 1        & 1        & 1 \\
  \alpha   & 0   & \alpha   & \alpha \\
  0        & \alpha^2 & \alpha^2 & \alpha^2 \\
  \alpha^2 & 1 & \alpha^2 & \alpha^2 \\
\end{matrix}\right)
\]

When the modified Brouwer-Zimmermann algorithm is applied to $A_4$, 
we start with $L=1$ and $U=4$. After enumerating one generator codewords we get $L=2$ and $U=3$, and after enumerating two generators codewords  we get $L=3$ and $U=3$, so the symplectic minimum distance is $3$.
In contrast, applying the same algorithm to $B_4$, 
we start with $L=1$ and $U=4$. After enumerating one generator codewords we get $L=2$ and $U=3$, and after enumerating two generators codewords  we get $L=3$ and $U=2$, so the symplectic minimum distance is $2$.

It is worth mentioning that the last two columns of  $A_4$ are not utilized in the diagonalization process. Generally, this could involve a number of columns of undetermined cardinality, which we refer to as the set of principal columns. Therefore, these columns may be rearranged to the first positions, and a new diagonalization can be performed solely with this set of principal columns.

This new matrix is denoted as $C_4$. Similarly, whenever a principal column contains exactly two different non-zero symbols, the addition of the corresponding rows is appended as a new row to  $C_4$, resulting in a new matrix, $D_4$.
This process is denoted as $D_4$ := \diagFcuatro($C_4$).

For the previous example, these matrices are 
\[
C_4=\left(\begin{matrix}
  1      & 1      & 1      & 1 \\
  \alpha & \alpha & \alpha & 0  \\
  0      & 0      & \alpha^2      & \alpha
\end{matrix}\right)
\]
and
\[
D_4=\left(\begin{matrix}
  1        & 1        & 1        & 1 \\
  \alpha   & \alpha   & \alpha   & 0  \\
  0        & 0        & \alpha^2      & \alpha \\
  \alpha^2 & \alpha^2 & \alpha^2 & 1 \\
\end{matrix}\right).
\]

Then, $B_4$ and $D_4$ are transformed back 
into matrices over $\mathbb{F}_2$ by reversing the process done previously.
Specifically, a row of the form $a+\alpha b\in\mathbb{F}_4^n$ is transformed into the row
$(a,b)\in\mathbb{F}_2^{2n}$.
The resulting matrices over $\mathbb{F}_2$ are denoted as $B_2$ and $D_2$.

In the previous example, these matrices are 

\[
B_2=\left(\begin{array}{@{}cccc|cccc@{}}
  1 & 1 & 1 & 1 & 0 & 0 & 0 & 0 \\
  0 & 0 & 0 & 0 & 1 & 0 & 1 & 1 \\
  0 & 1 & 1 & 1 & 0 & 1 & 1 & 1 \\
  1 & 1 & 1 & 1 & 1 & 0 & 1 & 1 \\
\end{array}\right)
\]
and
\[
D_2=\left(\begin{array}{@{}cccc|cccc@{}}
  1 & 1 & 1 & 1 & 0 & 0 & 0 & 0 \\
  0 & 0 & 0 & 0 & 1 & 1 & 1 & 0 \\
  0 & 0 & 1 & 0 & 0 & 0 & 1 & 1 \\
  1 & 1 & 1 & 1 & 1 & 1 & 1 & 0 \\
\end{array}\right).
\]

Subsequently, the modified Brouwer-Zimmermann algorithm 
is applied using the two generator matrices $B_2$ and $D_2$, 
obtaining the symplectic weight instead of the Hamming weight.

During this procedure, 
the second generator matrix $D_2$ contributes 
to adjusting the lower bound of the minimum distance 
only if a specific technical condition is met. 
Let us clarify this further. 
From the previous explanation of the matrices' construction, 
it is clear that we can analyze either $D_2$ or $D_4$. 
In this context, we will focus on $D_4$.

In our earlier discussion, it was shown
how to partition the rows of $B_4$ into packages, 
defining $\np{}(B_4)$ as the total number of these packages. 
Applying the same approach to $D_4$, 
$\np{}(D_4)$ represents the number of packages of the entire matrix $D_4$,
and $\Pnp{}(D_4)$ is the number of packages 
of the principal columns of the same matrix.

A key observation is that the diagonalization of $B_4$ separates it into two independent components: the principal columns, which have already been considered, and the remaining columns. As a result, it is possible that  $\np{}(B_4)$ might not be equal to $\np{}(D_4)$. Note that rows consisting entirely of zeros are excluded from forming any packages.

We have established that any codeword can be generated using at most one row from each package of $B_4$, meaning the number of generators $g$ considered in summations is at most $\np{}(B_4)$. Similarly, for $D_4$, the number of generators $g$ is constrained by $\np{}(D_4)$.

In conclusion,  when the number of generators is greater than or equal 
to $\np{}(D_4) - \Pnp{}(D_4)$, the matrix $D_2$ contributes 
with $g - \np{}(D_4) + \Pnp{}(D_4)+1$ 
to $L$, i.e., after fully enumerating all sums of $g$ generators we have
$$
L := g + 1 + \max\{ 0, g + 1 + \Pnp{}(D_4) - \np{}(D_4) \}.
$$

Consequently, based on our experimental results,
$D_2$ will only contribute when $k$ is relatively small. 
Section~\ref{performance} includes a specialized dataset,
\texttt{mat\_test4}, designed for matrices with small values of $k$. 
This dataset highlights the increased importance and effectiveness 
of the algorithm in such scenarios.
\begin{algorithm}[ht!]
  \caption{\ensuremath{\mbox{\sc \savedtwog}}}
  \label{alg:saved2g}
  \begin{algorithmic}[1]
    \REQUIRE A normalizer matrix $A$ of size $(n+k)\times 2n$ 
             of a quantum stabilizer code $Q$.
    \ENSURE  The minimum weight of $Q$.
    \medskip
    \STATE \textbf{Beginning of Algorithm}
    \STATE $A_4$ := \convertFFour(A):
    \STATE $B_4$ := \diagFcuatro($A_4$);
    \STATE $D_4$ := \diagFcuatro($B_4$);	
    \STATE $B_2$ := \convertFTwo($B_4$);
    \STATE $D_2$ := \convertFTwo($C_4$);
    \STATE d := \modifieds($\{B_2,D_2\}$);
    \RETURN d;
    \STATE \textbf{End of Algorithm}
  \end{algorithmic}
\end{algorithm}

% -----------------------------------------------------------------------------
\subsection{Algorithm \savedisom}
% -----------------------------------------------------------------------------

We essentially follow the methodology outlined by White~\cite{Grassl-white}.
The modified matrix $B$ is built from $A$ 
by extending every row of the latter,
denoted as $(a,b)\in\mathbb{F}_2^{2n}$, 
into the new row $(a,b,a+b)\in\mathbb{F}_2^{3n}$.
This mapping is denoted as \iso, since the Hamming weight of the image $(a,b,a+b)$ equals twice the symplectic weight of ($a,b)$.
The new matrix $B := \iso(A)$ indeed serves 
as the generator matrix of a linear code over $\mathbb{F}_2$.
Subsequently, the Hamming minimum distance 
of the linear code with the generator matrix provided by $B$ is computed.
To accomplish this task, the Modified Brouwer-Zimmermann algorithm
is used for computing the Hamming weight.
Finally, 
the true minimum distance of $Q$ is computed as the Hamming minimum
distance of $B$ divided by two.

\begin{algorithm}[ht!]
  \caption{\ensuremath{\mbox{\sc \savedisom}}}
  \label{alg:savedisom}
  \begin{algorithmic}[1]
    \REQUIRE A normalizer matrix $A$ of size $(n+k)\times 2n$ 
             of a quantum stabilizer code $Q$.
    \ENSURE  The minimum weight of $Q$.
    \medskip
    \STATE \textbf{Beginning of Algorithm}
    \STATE B := \iso(A);
    \STATE $d_1$ := \modified($B$);
    \STATE $d := \frac{d_1}{2}$;
    \RETURN d;
    \STATE \textbf{End of Algorithm}
  \end{algorithmic}
\end{algorithm}

% -----------------------------------------------------------------------------
\subsection{Availability of the implementations}
% -----------------------------------------------------------------------------

To achieve high performance, 
our algorithms have been implemented  with the C programming language,
which usually offers high speed when compiled to machine language 
(the native code of the CPU of the target computer).

To increase both the availability and simplicity of using our implementations,
we have compiled our C code to WebAssembly code
with the Emscripten compiler (release 3.1.61).
The WebAssembly~\cite{RossbergWebAssemblyCoreSpecification} language 
is a portable binary-code format,
with a higher level than machine languages.
Its main goal is to enable high-performance applications on web pages.
Hence, programs written in this language 
can be executed by the JavaScript interpreter included 
in most current browsers.
However, although the code can be executed in any browser, 
it is obviously much slower than native code.
Nevertheless, since the code can be executed in any browser,
it does not require any installation, just opening a web page.
This code is available at the following web page:

\url{gquintan.uji.es/symplectic_distance}

This page allows the user to compute the distance of quantum symplectic codes.
Since it is executed by the JavaScript interpreter of the browser, 
it is also executed in the local (user's computer) CPU.
Since JavaScript is single-threaded in browsers,
one drawback of this approach is that all output messages 
are only shown when all the computation has finished.

The input matrix $A$ is the normalizer matrix in extended format. 
If the quantum code is $Q$, 
the input matrix can be obtained in \magma{} with the following command:
\begin{lstlisting}
  A := NormalizerMatrix( Q: ExtendedFormat := true );
\end{lstlisting}

% =============================================================================
\section{Performance Analysis}
\label{performance}
% =============================================================================

In this study we assessed all the implementations on 
a server with 16 cores (AMD EPYC 7F52 at 2.0 GHz)
and a main memory of 512 GB.
The operating system is Ubuntu 20.04.6 LTS,
and the C compiler is GNU gcc (Ubuntu 9.4.0-1ubuntu1~20.04.2) 9.4.0.

Several other computers were assessed,
obtaining similar results with our implementations.
The results on these computers are not reported
since \magma{} was not installed on them 
(since it is a licensed software),
and therefore no comparison could be performed.

The implementations assessed in this work are the following ones:

\begin{itemize}

\item
\magma{}~\cite{Magma}:
It is a licensed software package designed for computations in algebra,
algebraic combinatorics, algebraic geometry, etc.
Version V2.26-10 was employed in our experiments.
The implementation without AVX vectorization was employed
since it was faster than the vectorized versions
due to the short length of the vectors being processed.

\item
\savedoneg:
This implementation of ours uses one $\Gamma$ matrix
by using a diagonalization over $\mathbb{F}_2$.
No vectorization was employed for the sake of a fair comparison with \magma{}.

\item
\savedtwog:
This implementation of ours uses two $\Gamma$ matrices
by using a diagonalization over $\mathbb{F}_4$.
No vectorization was employed for the same reason.

\item
\savedisom:
This implementation of ours uses the isometry method described above.
No vectorization was employed for the same reason.

\end{itemize}

All plots included in this study are of two types:
Some plots show the times, and therefore lower is better.
On the other hand, other plots show the speedups of the new implementations
when compared to \magma{}, and therefore higher is better
for the new implementations.
The speedup of an implementation is computed
as the time of \magma{} divided by the time of that implementation,
and therefore it is the number of times that the latter faster than \magma{}.

% -----------------------------------------------------------------------------
% Overall results.
% -----------------------------------------------------------------------------

In order to assess all the implementations,
about two thousand normalizer matrices in extended format of dimension $K\times N$ with $K=n+k$ and $N=2n$ for stabilizer codes $[\![n,k,d]\!]_2$ were processed,
all of them generated randomly.
These matrices were grouped
into the following four different tests or datasets:

\begin{itemize}

\item
\texttt{mat\_test1}:
It comprises $286$ small matrices generated randomly.
The maximum number of elements of the matrices in this dataset was $3\,552$.
The minimum, average, and maximum ratio $K/N$ of these matrices
(of dimension $K \times N$)
was $0.519$, $0.673$, and $0.958$, respectively.
This dataset was employed
to check that the symplectic distances computed by
our new implementations were the same as those of \magma{}.
Since the matrices were small, the computational time was very small,
and therefore performances are not reported.

\item
\texttt{mat\_test2}:
It comprises $1650$ small matrices generated randomly.
The maximum number of elements of the matrices in this dataset was $3\,384$.
The minimum, average, and maximum ratio $K/N$ of these matrices
(of dimension $K \times N$)
was $0.519$, $0.677$, and $0.958$, respectively.
This dataset was employed to check the symplectic distances.
Since the matrices were small, performances are not reported.

\item
\texttt{mat\_test3}:
It comprises $300$ medium and large matrices generated randomly.
The maximum number of elements of the matrices in this dataset was $12\,416$.
The minimum, average, and maximum ratio $K/N$ of these matrices
(of dimension $K \times N$)
was $0.586$, $0.654$, and $0.758$, respectively.
Because of this,
usually only one $\Gamma$ matrix contributed
to the lower bound in the Brouwer-Zimmermann algorithm.
As this dataset and the next one contained larger matrices
(usually with a larger computational cost),
they were employed to assess performances.
Nevertheless, symplectic distances were also checked.

Matrices of this dataset in which \magma{} took less than $1$ second
when using one core were discarded, thus keeping $221$ matrices in total
with significant computational times.
These remaining matrices were classified according
to their computational time in \magma{} when using one core
into the following subtests or subdatasets:

  \begin{itemize}
  \item Subtest $a$: 
        It contains those matrices in which \magma{} took $[1,10)$ seconds.
  \item Subtest $b$: 
        It contains those matrices in which \magma{} took $[10,100)$ seconds.
  \item Subtest $c$: 
        It contains those matrices in which \magma{} took $[100,1\,000)$ seconds.
  \item Subtest $d$: 
        It contains those matrices in which \magma{} took $[1\,000,10\,000)$ seconds.
  \item Subtest $e$: 
        It contains those matrices in which \magma{} took $10\,000$ seconds or more.
  \end{itemize}

\item
\texttt{mat\_test4}:
It comprises $11$ medium and large matrices
especially designed to benefit algorithms that employ two $\Gamma$
matrices.
The maximum number of elements of the matrices in this dataset was $9\,100$.
The minimum, average, and maximum ratio $K/N$ of these matrices
(of dimension $K \times N$)
was $0.538$, $0.542$, and $0.545$, respectively.
As before, symplectic distances were also checked.

\end{itemize}

% -----------------------------------------------------------------------------
\subsection{Performances on one core}
% -----------------------------------------------------------------------------

% -----------------------------------------------------------------------------
% mat_test3 (one gamma)
% -----------------------------------------------------------------------------

% -----------------------------------
% Box plots of matrix test 3.
% -----------------------------------

\begin{figure}[ht!]
\tfvspace
\begin{center}
\begin{tabular}{cc}
\includegraphics[width=0.45\textwidth]{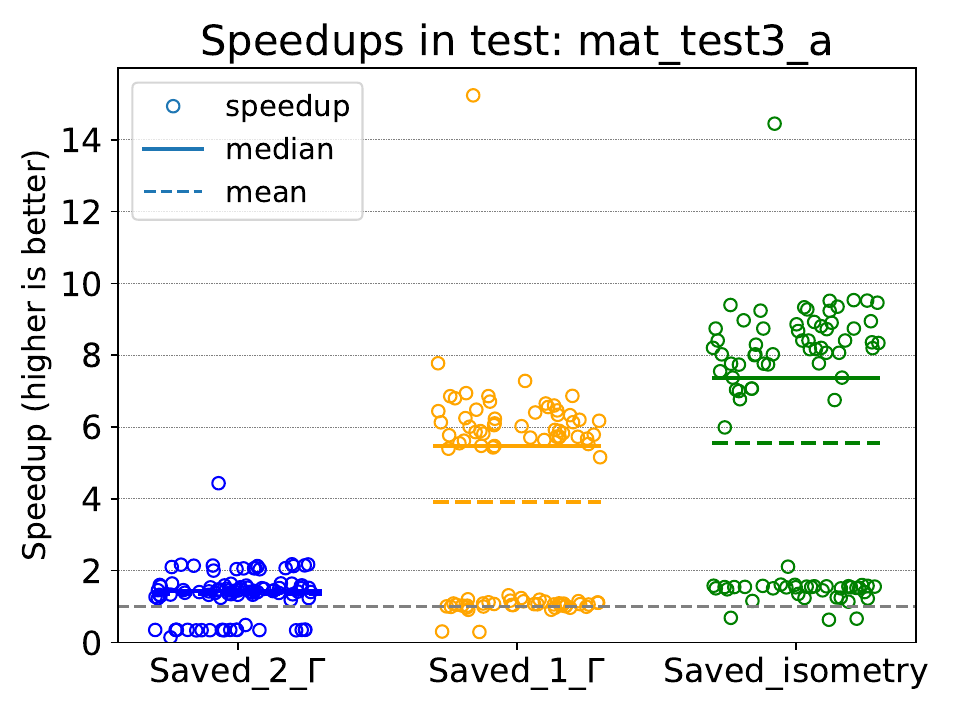}
&
\includegraphics[width=0.45\textwidth]{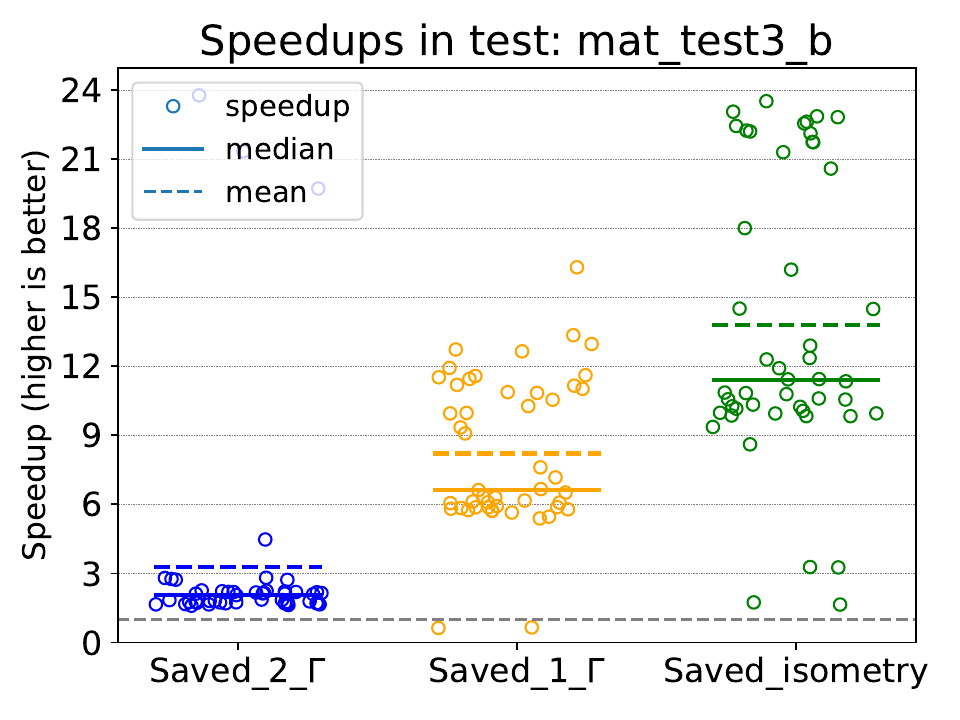}
\\
\includegraphics[width=0.45\textwidth]{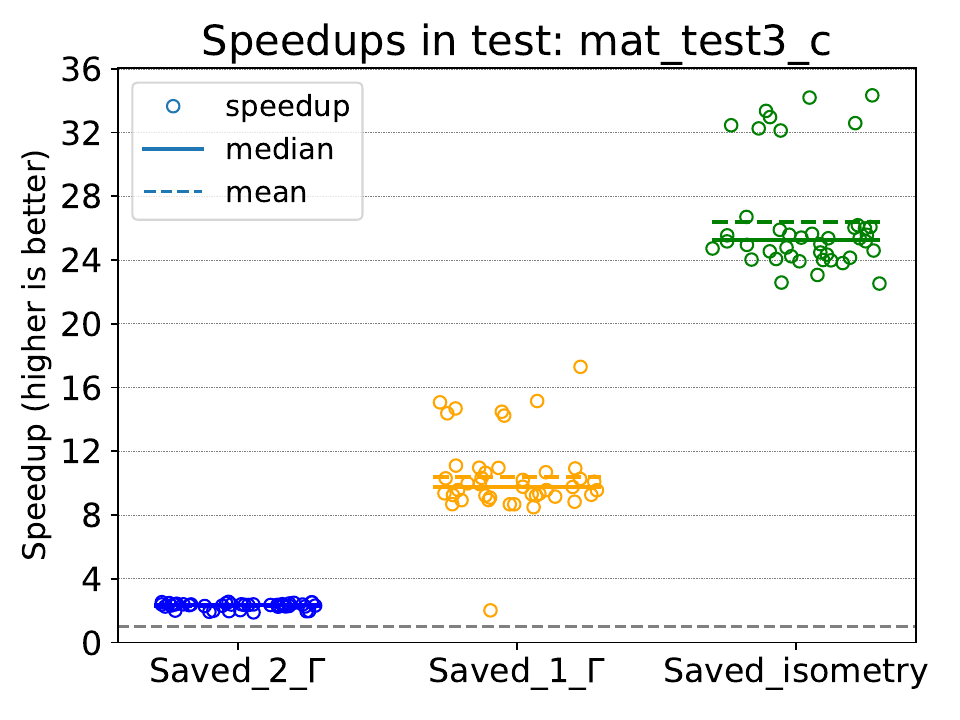}
&
\includegraphics[width=0.45\textwidth]{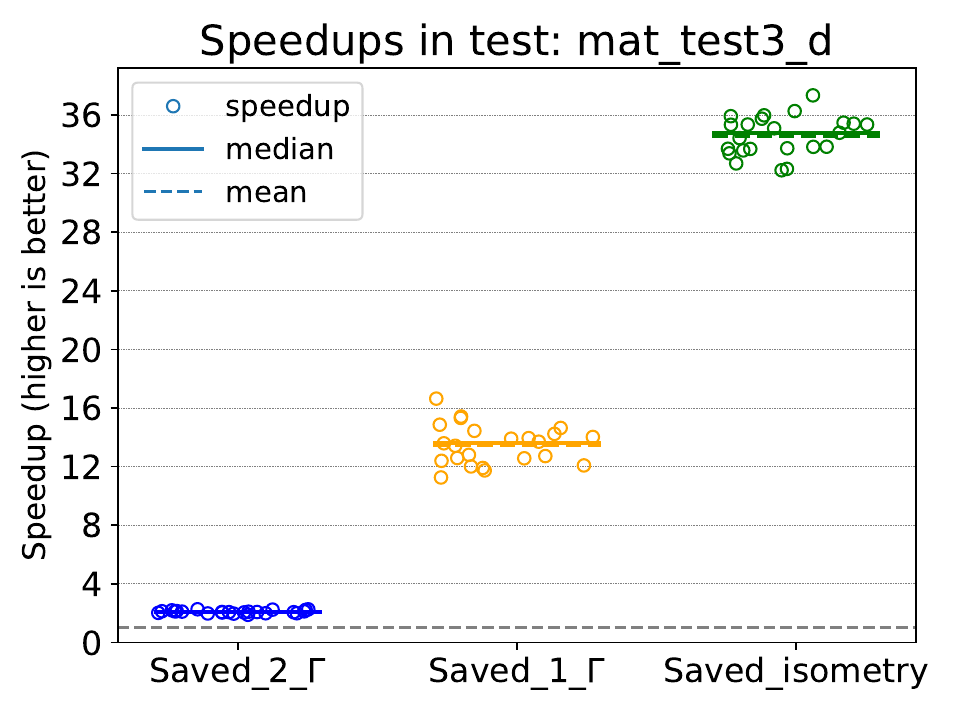}
\\
\multicolumn{2}{c}{
\includegraphics[width=0.45\textwidth]{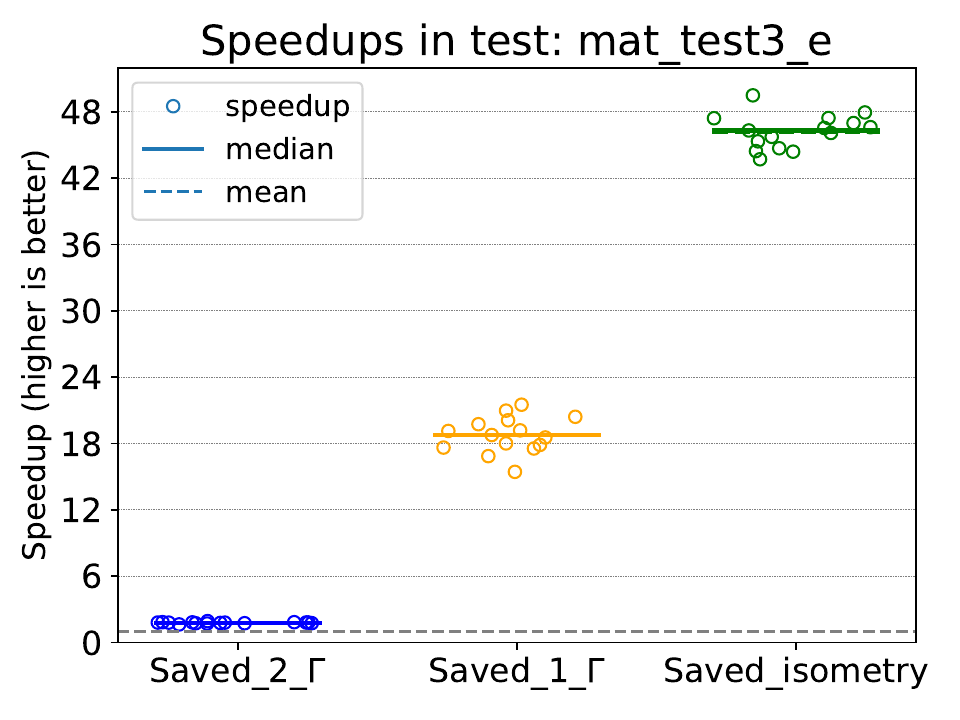}
} \\
\end{tabular}
\end{center}
\bfvspace
\caption{Speedups of our new implementations with respect to \magma{}
for all matrices in the \texttt{mat\_test3} dataset.
}
\label{fig:swa_matrix_test3}
\end{figure}

This subsection reports results on the \texttt{mat\_test3}
and \texttt{mat\_test4} datasets,
since their computational times are much longer than the first two
datasets.

When assessing all implementations for the \texttt{mat\_test3} dataset
on one core,
our new implementations \savedtwog{}, \savedoneg{}, and \savedisom{}
outperformed \magma{} in
$200$ cases out of $221$ ($90.5$ \%),
$209$ cases out of $221$ ($94.6$ \%), and
$218$ cases out of $221$ ($98.6$ \%), respectively.
The few cases in which \magma{} was faster
belonged to the subtests \textit{a} and \textit{b}
(and most of them to the first one).
Recall that subtest \textit{a} comprises
all cases with computational times in $[1.0, 10.0)$
and subtest \textit{b} comprises
all cases with computational times in $[10.0, 100.0)$.

Figure~\ref{fig:swa_matrix_test3} shows a comparison of \magma{} and
our implementations for all matrices in the \texttt{mat\_test3} dataset.
To compare both \magma{} and our new implementations, 
this figure shows speedups.
A circle $\circ$ represent the data point for the speedup for a matrix.
For example, if the vertical coordinate of one of our implementations 
is $10$, it means that it is $10$ times as fast as \magma{}.
To avoid many symbols overlapping in one place,
a random small value has been added to the horizontal coordinate.
The continuous line and the dashed line represent 
the median and the mean, respectively, of all the speedups of one implementation.

As this figure shows,
the speedups of \savedtwog{} are usually larger than one,
thus being faster than \magma{} in most cases.
On the other side,
the speedups of \savedoneg{} and \savedisom{} are indeed remarkable.
The median speedups of
these two implementations with respect to \magma{}
are usually between around $5$ and around $45$.
Therefore, the computational times of our two implementations
are between around $5$ and around $45$ times smaller than \magma{} in median.
It is also important to note that
the speedups obtained grow as the computational times grow.
Recall that each subtest (shown in one plot of that figure)
requires a higher (around an order of magnitude) computational time 
than the previous one.

% -----------------------------------
% Bar plots of matrix test 3.
% -----------------------------------

% --------------------
% Test 3a
% --------------------
\begin{figure}[ht!]
\tfvspace
\begin{center}
\begin{tabular}{cc}
\includegraphics[width=0.45\textwidth]{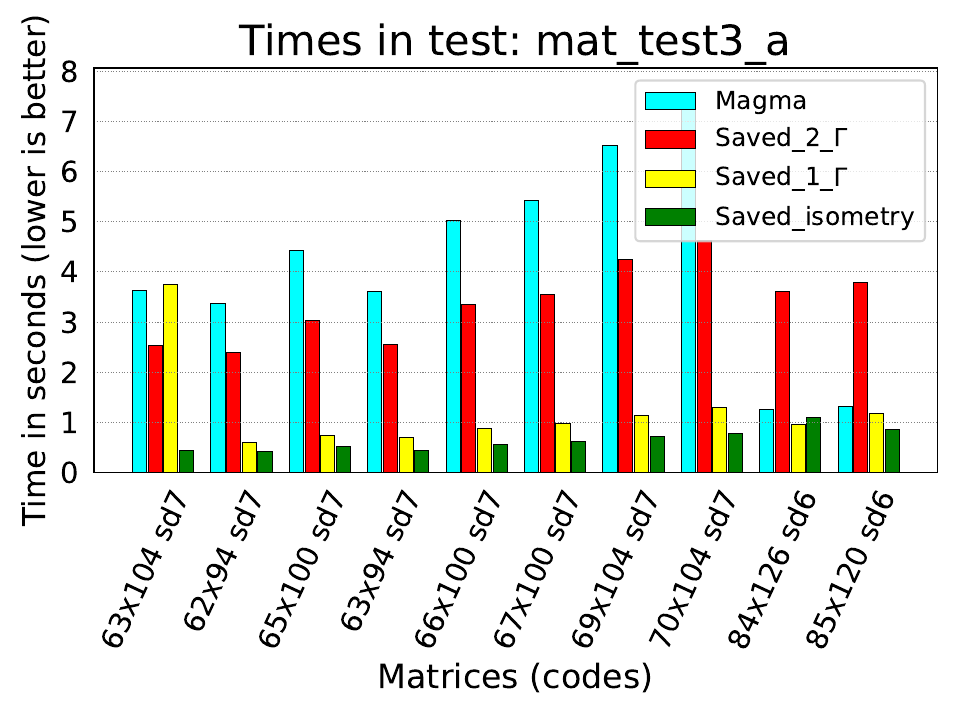}
&
\includegraphics[width=0.45\textwidth]{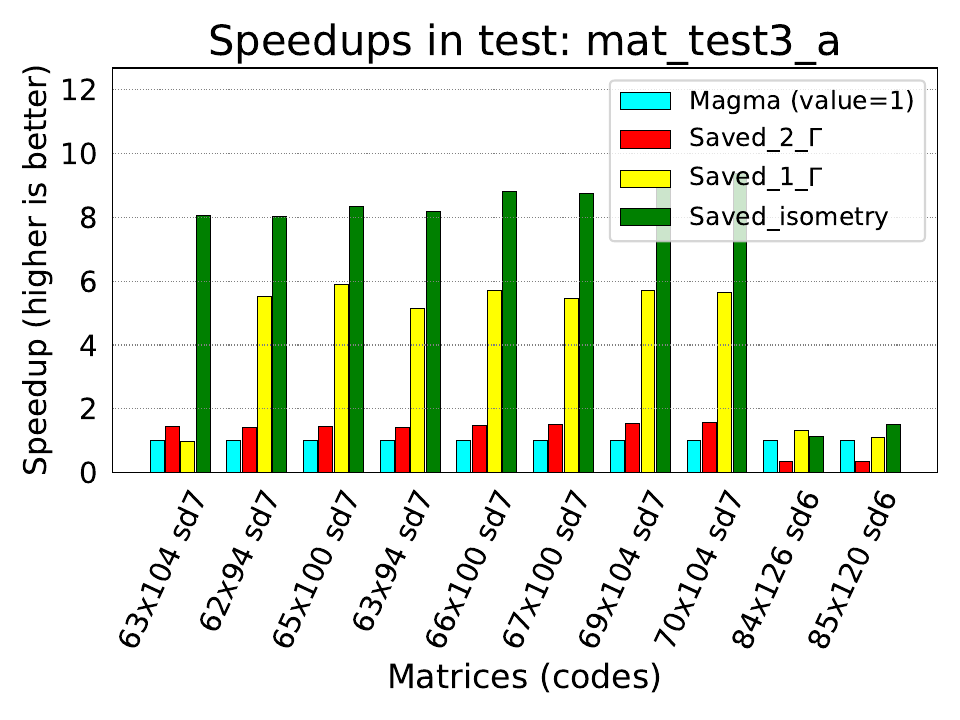}
\\
\end{tabular}
\end{center}
\bfvspace
\caption{Time in seconds (left) and speedups (right)
for several matrices.
In these plots \magma{} time is $[1, 10)$.
The horizontal axis shows the matrices assessed
with their dimensions ($K \times N$) and 
their symplectic distance (\textit{sd}).
}
\label{fig:matrix_test3_a}
\end{figure}

% --------------------
% Test 3b
% --------------------
\begin{figure}[ht!]
\tfvspace
\begin{center}
\begin{tabular}{cc}
\includegraphics[width=0.45\textwidth]{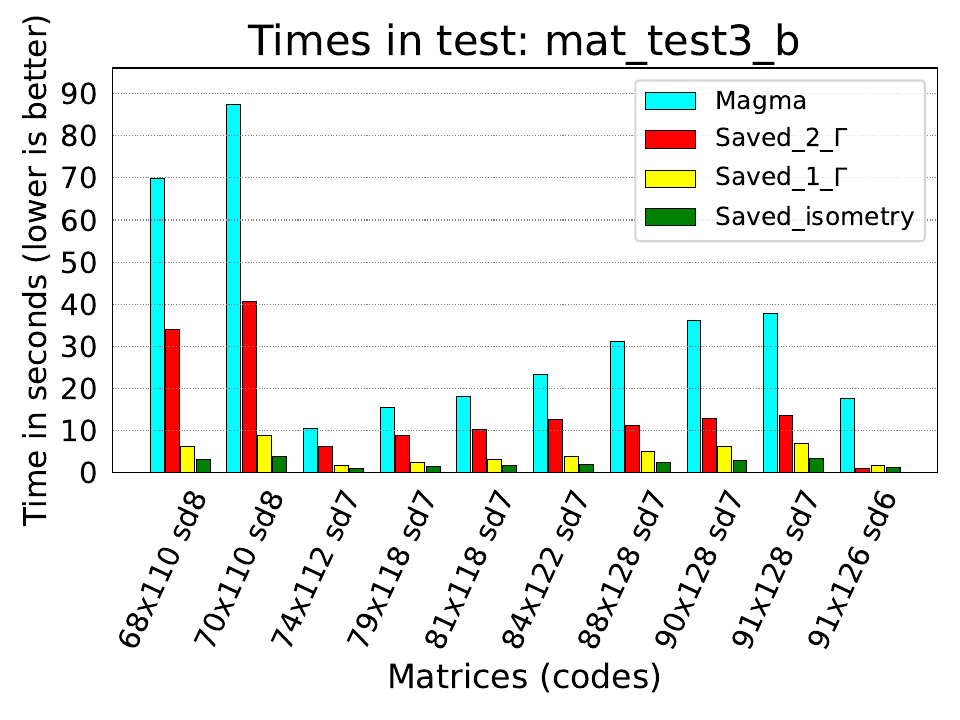}
&
\includegraphics[width=0.45\textwidth]{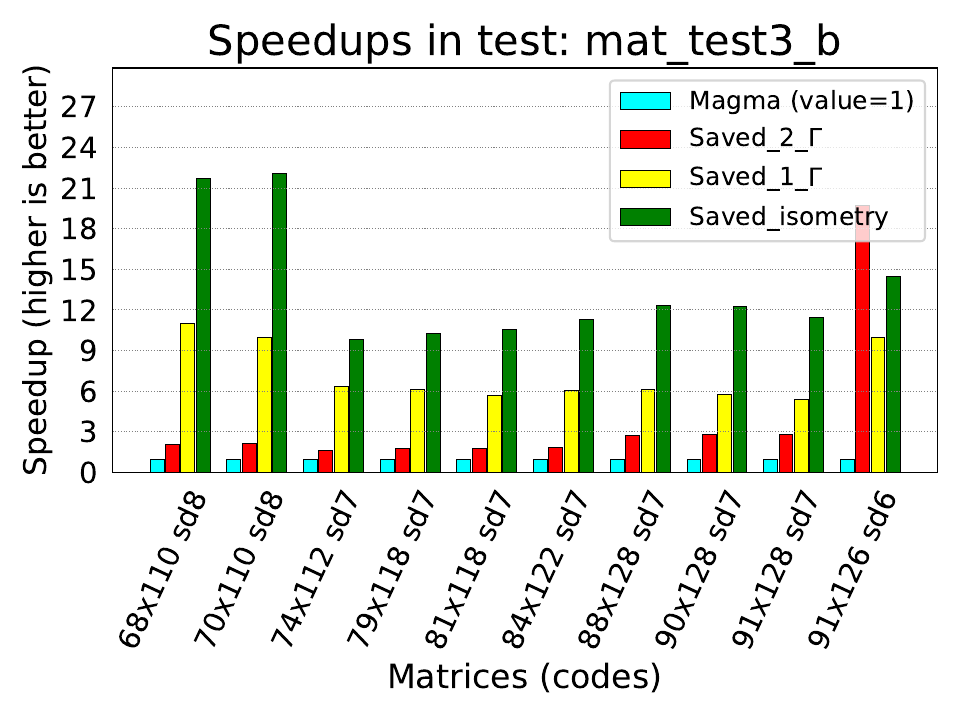}
\\
\end{tabular}
\end{center}
\bfvspace
\caption{Time in seconds (left) and speedups (right)
for several matrices.
In these plots \magma{} time is $[10, 100)$.
The horizontal axis shows the matrices assessed
with their dimensions ($K \times N$) and 
their symplectic distance (\textit{sd}).
}
\label{fig:matrix_test3_b}
\end{figure}

% --------------------
% Test 3c
% --------------------
\begin{figure}[ht!]
\tfvspace
\begin{center}
\begin{tabular}{cc}
\includegraphics[width=0.45\textwidth]{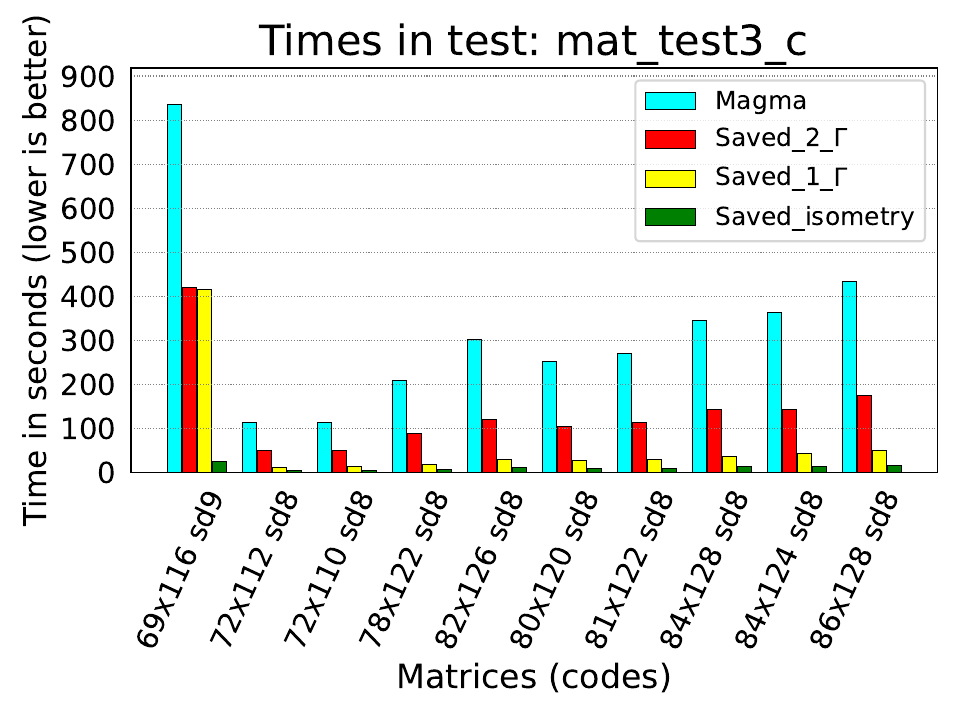}
&
\includegraphics[width=0.45\textwidth]{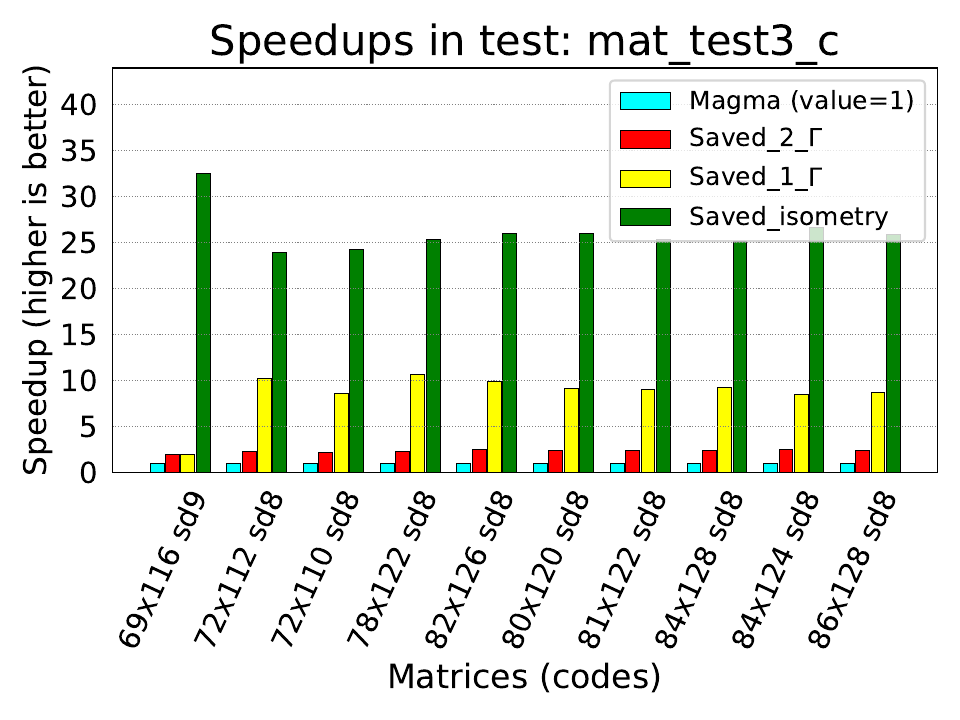}
\\
\end{tabular}
\end{center}
\bfvspace
\caption{Time in seconds (left) and speedups (right)
for several matrices.
In these plots \magma{} time is $[100, 1000)$.
The horizontal axis shows the matrices assessed
with their dimensions ($K \times N$) 
and their symplectic distance (\textit{sd}).
}
\label{fig:matrix_test3_c}
\end{figure}

% --------------------
% Test 3d
% --------------------
\begin{figure}[ht!]
\tfvspace
\begin{center}
\begin{tabular}{cc}
\includegraphics[width=0.45\textwidth]{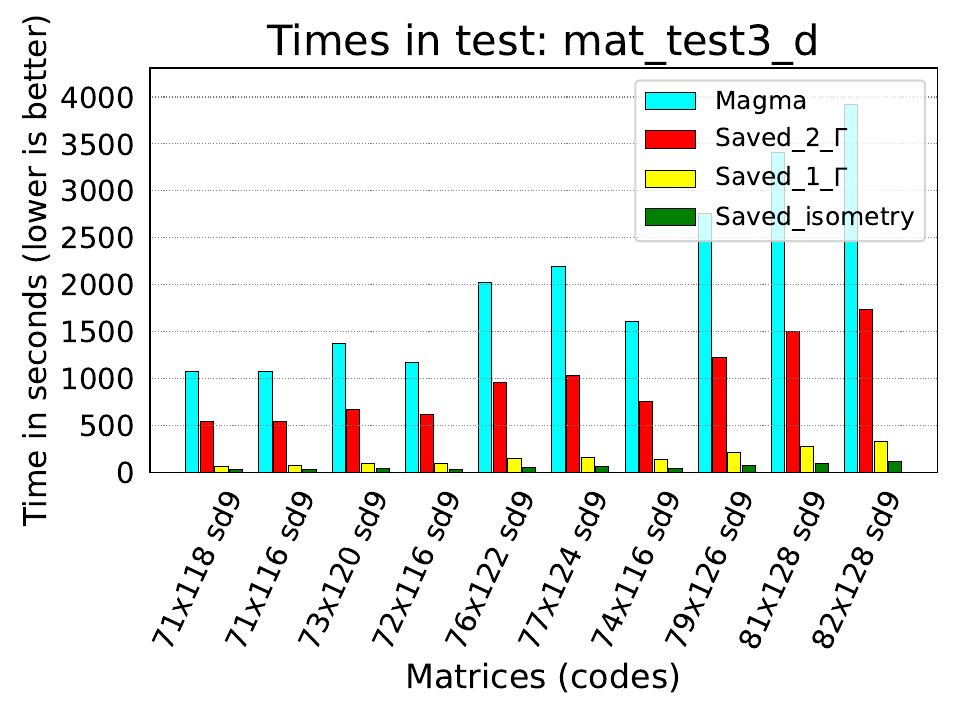}
&
\includegraphics[width=0.45\textwidth]{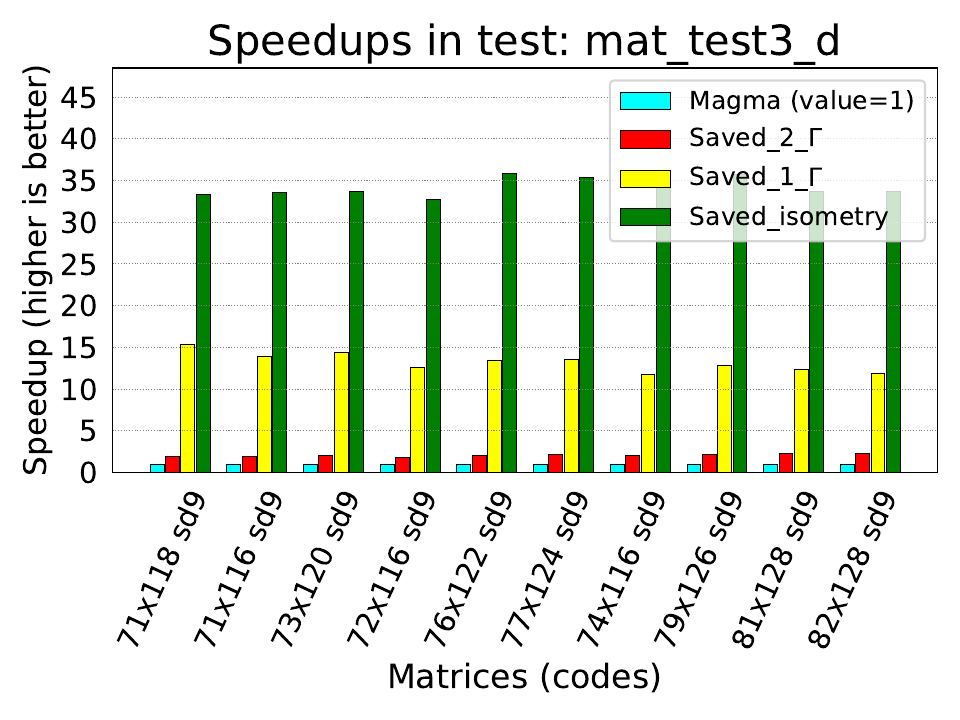}
\\
\end{tabular}
\end{center}
\bfvspace
\caption{Time in seconds (left) and speedups (right)
for several matrices.
In these plots \magma{} time is $[1000, 10000)$.
The horizontal axis shows the matrices assessed
with their dimensions ($K \times N$) and 
their symplectic distance (\textit{sd}).
}
\label{fig:matrix_test3_d}
\end{figure}

% --------------------
% Test 3e
% --------------------
\begin{figure}[ht!]
\tfvspace
\begin{center}
\begin{tabular}{cc}
\includegraphics[width=0.45\textwidth]{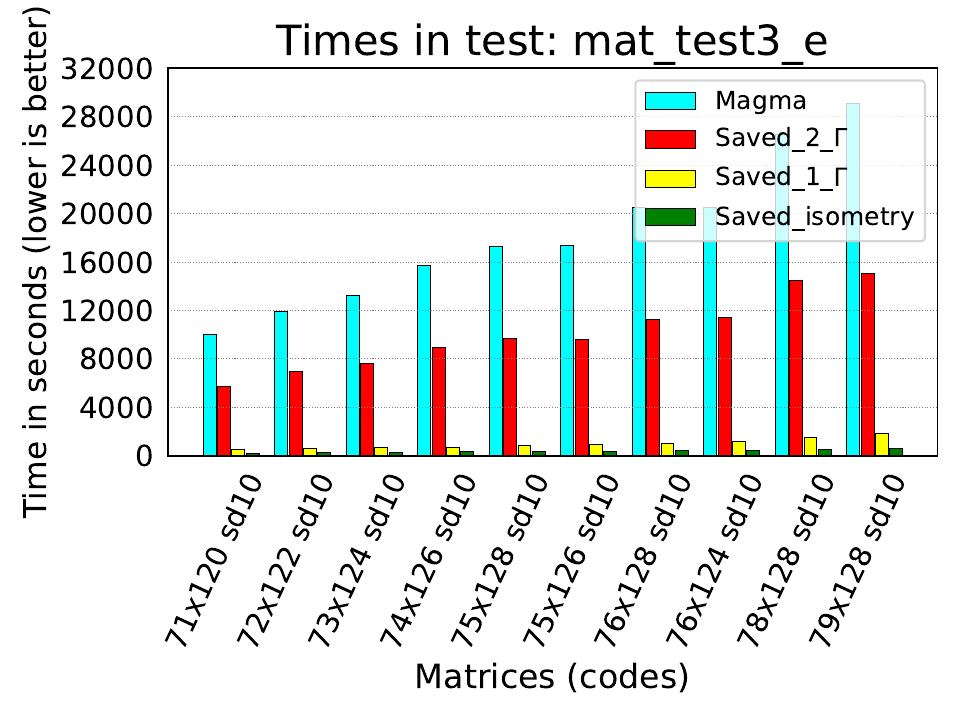}
&
\includegraphics[width=0.45\textwidth]{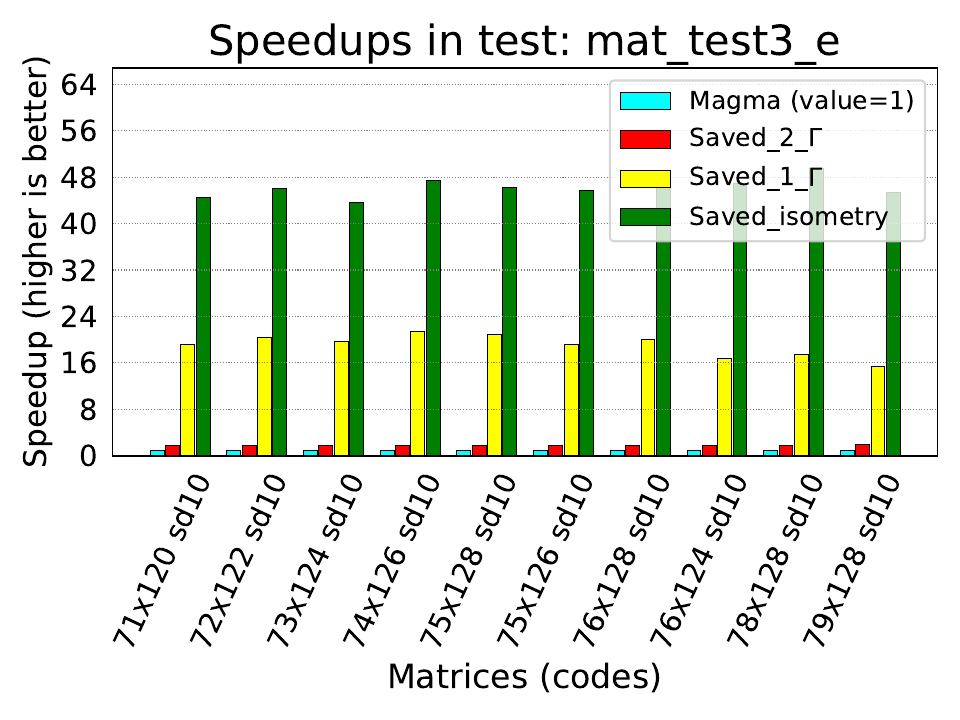}
\\
\end{tabular}
\end{center}
\bfvspace
\caption{Time in seconds (left) and speedups (right)
for several matrices.
In these plots \magma{} time is $\ge 10000$ s.
The horizontal axis shows the matrices assessed
with their dimensions ($K \times N$) and 
their symplectic distance (\textit{sd}).
}
\label{fig:matrix_test3_e}
\end{figure}

To analyze the previous results in more detail,
Figures~\ref{fig:matrix_test3_a},
\ref{fig:matrix_test3_b},
\ref{fig:matrix_test3_c},
\ref{fig:matrix_test3_d}, and
\ref{fig:matrix_test3_e}
show a comparison of \magma{} and our implementations
for a set of random samples extracted
from each subset of the \texttt{mat\_test3} dataset.
Each figure shows both the time in seconds (left plots)
and the speedups (right plots)
for the sample of the corresponding subset.
The horizontal axis shows the matrix dimensions ($K \times N$)
and the symplectic distance (\textit{sd}) in both types of plots.

As can be seen,
our \savedoneg{} and \savedtwog{} implementations
outperform \magma{} in most cases.
It is also important to note how the speedups grow significantly
as the computational time of \magma{} grows.

% -----------------------------------------------------------------------------
% mat_test4 (two gammas)
% -----------------------------------------------------------------------------

\begin{figure}[ht!]
\tfvspace
\begin{center}
\begin{tabular}{cc}
\includegraphics[width=0.45\textwidth]{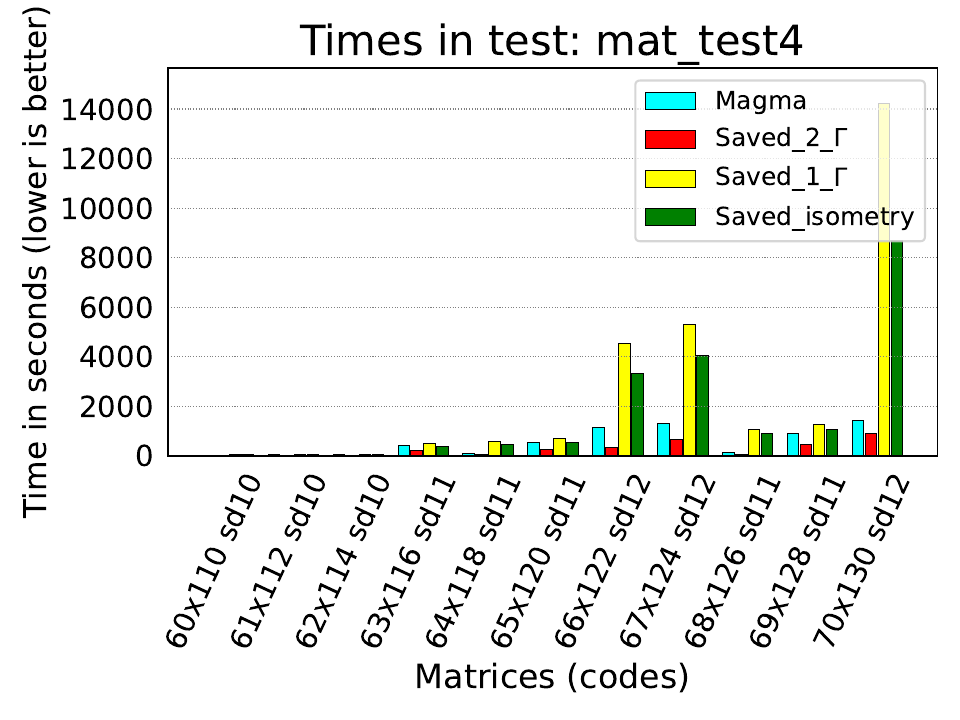}
&
\includegraphics[width=0.45\textwidth]{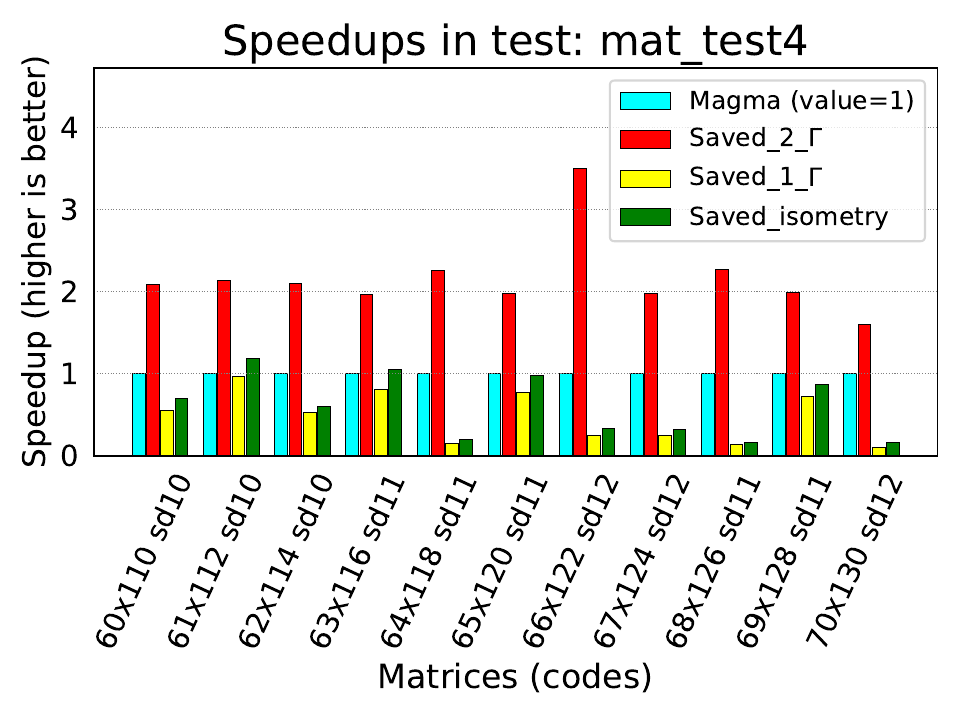}
\\
\end{tabular}
\end{center}
\bfvspace
\caption{Time in seconds (left) and speedups (right)
for several matrices.
The horizontal axis shows the matrices assessed
with their dimensions ($K \times N$) and 
their symplectic distance (\textit{sd}).
}
\label{fig:matrix_test4}
\end{figure}

Figure~\ref{fig:matrix_test4}
shows a comparison of \magma{} and our implementations
on the \texttt{mat\_test4} dataset.
This dataset contains matrices specially searched so that
two $\Gamma$ matrices would accelerate the time
by contributing to the lower bound in the Brouwer-Zimmermann algorithm.
As shown, \savedtwog{} is usually around two times as fast as \magma{}.

% -----------------------------------------------------------------------------
\subsection{Parallel performances}
% -----------------------------------------------------------------------------

\begin{table}[ht!]
\caption{Time in seconds of \magma{} and our implementations
on one matrix of every subset
when using several cores.}
\begin{center}
\begin{tabular}{lcrrrrr} \hline
                 & \multicolumn{1}{c}{No.}
                 & \multicolumn{1}{c}{subtest a}
                 & \multicolumn{1}{c}{subtest b}
                 & \multicolumn{1}{c}{subtest c}
                 & \multicolumn{1}{c}{subtest d}
                 & \multicolumn{1}{c}{subtest e} \\
  \multicolumn{1}{c}{Implementation} & \multicolumn{1}{c}{cores}
                 & \multicolumn{1}{c}{\texttt{mat30005}}
                 & \multicolumn{1}{c}{\texttt{mat30013}}
                 & \multicolumn{1}{c}{\texttt{mat30015}}
                 & \multicolumn{1}{c}{\texttt{mat30042}}
                 & \multicolumn{1}{c}{\texttt{mat30020}} \\ \hline\hline
  \magma{}     &  1 &  1.63 & 46.55 &    583.64 & 1\,076.66 & 10\,012.45 \\
               &  4 &  3.15 & 80.69 & 1\,294.61 & 2\,756.03 & 33\,722.99 \\
  \hline
  \savedtwog{} &  1 & 1.32 & 21.90 & 300.78 & 524.16 & 5\,769.85 \\
               &  4 & 0.88 &  6.75 &  78.93 & 135.90 & 1\,450.25 \\
               &  8 & 0.79 &  4.21 &  44.16 &  75.40 &    804.33 \\
  \hline
  \savedoneg{} &  1 & 0.28 &  3.84 &  38.33 &  68.16 &    518.73 \\
               &  4 & 0.15 &  1.07 &   9.43 &  17.69 &    132.88 \\
               &  8 & 0.13 &  0.66 &   5.28 &   9.91 &     70.80 \\
  \hline
  \savedisom{} &  1 & 0.22 &  2.07 &  17.48 &  30.98 &    224.84 \\
               &  4 & 0.16 &  0.73 &   4.65 &   8.43 &     57.98 \\
               &  8 & 0.16 &  0.52 &   2.64 &   4.78 &     30.27 \\
  \hline\hline
\end{tabular}
\end{center}
%%%% \bfvspace
\label{tab:speedups}
\end{table}

Table~\ref{tab:speedups}
compares the computational times of \magma{} and our implementations
with respect to the number of cores being employed.
For every subset of \texttt{mat\_test3},
the first matrix of the sample employed above was assessed.
The \magma{} implementation has only been assessed on 1 and 4 cores
since their performances decrease as the number of cores increases.
As said before,
the performances on one core of
our \savedoneg{} and \savedisom{} are remarkable.
Moreover,
unlike \magma{} (where using more than one core results often in a slow-down), our new implementations significantly accelerate their
performances when using several cores,
thus taking full advantage of modern shared-memory parallel architectures.
For example, a computation (\texttt{mat30020})
that takes more than two hours and a half in \magma{}
(when using one core and much more when using four cores)
takes around half a minute in one of our new implementations.

Note the good scalability of our new implementations
since in our implementations
the ratio of the times on $1$ core and $4$ cores
is usually very close to four,
and that the ratio of the times on $1$ core and $8$ cores
is usually very close to eight.

% =============================================================================
\section{Conclusions}
\label{conclusions}
% =============================================================================

The symplectic distance of a stabilizer quantum code 
is a very important feature since it determines the number of errors
that can be detected and corrected.
This work presents three new fast implementations
for computing the symplectic distance.
Our new implementations are based 
on recent fast implementations of the Brouwer-Zimmermann algorithm.
Our experimental study included several thousands of matrices.
It shows that our new implementations are much faster 
than current state-of-the-art licensed implementations
on single-core processors,
multicore processors,
and share-memory multiprocessors.
In the most computationally-demanding cases,
the performance gain in the computational time of our new implementations
is usually larger than one order of magnitude.
The largest performance gain in the computational time 
of our new implementations
observed in the experimental study was around $45$ times as fast.
The scalability of our new implementations 
on shared-memory parallel architectures is also very good.

% =============================================================================
\section*{Contributor role statement}
% =============================================================================

\textbf{Hernando}:
Conceptualization,
Methodology,
Software,
Validation,
Formal analysis,
Investigation,
%%%% Resources,
Data Curation,
Writing - Original Draft,
%%%% Writing - Review \& Editing,
Visualization.
%%%% Supervision,
%%%% Project administration,
%%%% Funding acquisition.
%
%
\textbf{Quintana-Ortí}:
Conceptualization,
Methodology,
Software,
Validation,
Formal analysis,
Investigation,
%%%% Resources,
Data Curation,
Writing - Original Draft,
%%%% Writing - Review \& Editing,
Visualization.
%%%% Supervision,
%%%% Project administration,
%%%% Funding acquisition.
%
%
\textbf{Grassl}:
Conceptualization,
%%%% Methodology.
%%%% Software,
Validation,
%%%% Formal analysis,
%%%% Investigation,
%%%% Resources,
%%%% Data Curation,
Writing - Original Draft.
\section*{Acknowledgements}
% =============================================================================

F. Hernando was partially funded by MCIN/AEI/10.13039/501100011033, by 
``ERDF: A way of making Europe'' and by ``European Union NextGeneration EU/PRTR'' 
Grants PID2022-138906NB-C22 and TED2021-130358B-I00, 
as well as by Universitat Jaume I, Grants UJI-B2021-02 and GACUJIMB-2023-03.

G. Quintana-Ortí was supported by
the Spanish Ministry of Science, Innovation and Universities
under Grant RTI2018-098156-B-C54 co-ﬁnanced with FEDER funds.

M. Grassl would like to thank for the hospitality 
during his visit of the University Jaume I.
The `International Centre for Theory of Quantum Technologies’ project 
(contract no. 2018/MAB/5) is carried out 
within the International Research Agendas Programme of the Foundation 
for Polish Science co-financed by the European Union 
from the funds of the Smart Growth Operational Programme, 
axis IV: Increasing the research potential (Measure 4.3).

% =============================================================================
% Bibliography
% =============================================================================

\bibliography{bibliography}
\bibliographystyle{amsplain}

% =============================================================================
\end{document}